\documentclass[5p,npa,10pt]{elsarticle}
\usepackage{graphics}
\usepackage{rotating}
\usepackage{mathrsfs}
\usepackage{amsmath}
\usepackage{lscape}
\usepackage[usenames,dvipsnames]{color}
\usepackage{tikz,ifthen,calc}
\usetikzlibrary{snakes,shapes}
\usetikzlibrary{patterns}
\usetikzlibrary{plotmarks}
\usetikzlibrary{calc}
\usepackage{pgfplots}
\usepackage[utf8]{inputenc}
\usepackage{graphicx} 
\usepackage{amsmath,amssymb}
\usepackage{natbib}
\usepackage[utf8]{inputenc}
\usepackage{graphicx} 
\usepackage{amsmath,amssymb}
\usepackage{multirow}

\DeclareUnicodeCharacter{00A0}{ }
\begin{document}
\title{Empirical pairing gaps, shell effects, and di-neutron spatial correlation in neutron-rich nuclei}%

\author{S. A. Changizi}%
\author{Chong Qi}%
\ead{chongq@kth.se}
\author{R. Wyss}
\address{Department of Physics, Royal Institute of Technology (KTH), SE-10691 Stockholm, Sweden}

\begin{abstract} The empirical pairing gaps derived from
four different odd-even mass staggering formulas are compared. By performing single-$j$ shell and multi-shell seniority model calculations  as well as by using the standard HFB approach  
with Skyrme force we show
that the simplest three-point formula $\Delta_C^
{(3)}(N)=\frac{1}{2}\left[B(N,Z)+B(N-2,Z)-2B(N-1,Z)\right]$ can provide a good
measure of the neutron pairing gap in even-$N$ nuclei. It removes to a large extent
the contribution from the nuclear mean field as well as contributions from
shell structure details.  It is also less contaminated by the Wigner effect for nuclei
around $N=Z$. We also show that the strength of $\Delta^ {(3)}_C(N)$ can serve as a good indication of the two-particle spatial correlation in the nucleus of concern and that 
the weakening of $\Delta^ {(3)}_C(N)$ in some neutron-rich 
nuclei indicates that the di-neutron correlation itself is weak in these nuclei. 
\end{abstract}

\maketitle

The occurrence of a systematic odd-even staggering (OES) of the nuclear binding 
energy has long been identified in nuclear physics, which  is associated with the
pairing correlation \cite{ring2004nuclear,bohr1998nuclear}. It plays an important role in many 
nuclear phenomena and is the  dominant many-body correlation beyond the
nuclear mean field. Yet, in spite of the many efforts performed in the study
of pairing correlations, there are still features which may be
induced by the pairing interaction that are not well understood \cite{FiftyYears2013,PhysRevLett.109.032501}. 
In particular, this is the case in neutron-rich nuclei, where the study of 
effects induced by pairing may shed light on the understanding of various 
exotic phenomena (see, e.g., Refs. \cite{
PhysRevC.88.034314,Anguiano2013,PhysRevC.88.054308,
PhysRevC.85.034317}).

The simplest expression one can use to extract the empirical pairing gap from the OES of the
binding energy is the three-point formula 
\cite{bohr1998nuclear,PhysRevLett.81.3599}, which for systems with even 
neutrons acquires the form~\cite{bohr1998nuclear}
\begin{multline}
\label{eq:3point}
\Delta^ {(3)}(N)=-\frac{1}{2}\left[B(N-1,Z)+B(N+1,Z)-2B(N,Z)\right]\\
=-\frac{1}{2}[S_n(N+1,Z)-S_n(N,Z)]
\end{multline} 
where $B$ is the (positive) binding energy and $S_n$ is the one-neutron 
separation energy.  The proton pairing gap can be defined in a similar way. 
The above formula indicates that $\Delta^ {(3)}(N)$ measures the additional 
binding gain by the last neutron in the even-$N$ system relative to the odd 
system with one more neutron. However, 
besides pairing, a number of other mechanisms may contribute to the OES 
\cite{PhysRevLett.81.3599,PhysRevC.88.064329,Friedman2009,PhysRevC.60.051301}. This includes
effects induced by the mean field in deformed nuclei (or the Kramers 
degeneracy) and the contribution from the diagonal
interaction matrix elements of the two-body force. 

As discussed in detail in Refs. 
\cite{PhysRevLett.81.3599,PhysRevC.63.024308} by Satula and co-workers, the contribution from the quickly varying single-particle structure of the mean field to the empirical  pairing gap is minimized in odd-mass systems.
In even systems where the last neutrons occupy different orbitals the single-particle energy
contributes substantially to $\Delta^ {(3)}(N)$\cite{PhysRevLett.81.3599,PhysRevC.63.024308}.

Alternatively, there is another version of the three-point formula written as
\begin{multline}
\label{eq:3pointC}
\Delta^ {(3)}_{C}(N)=\frac{1}{2}[S_n(N,Z)-S_n(N-1,Z)]\\
=\frac{1}{2}\left[B(N,Z)+B(N-2,Z)-2B(N-1,Z)\right] \\
=\frac{1}{2}[S_{2n}(N,Z)-2S_n(N-1,Z)],
\end{multline} 
which actually corresponds to $\Delta^ {(3)}$ for the case of odd nuclei 
\cite{PhysRevLett.81.3599}.  $\Delta^ {(3)}(N-1)$ is smaller than $\Delta^ {(3)}(N)$ in most cases. It is often stated that $\Delta^ {(3)}(N-1)$ measures the pairing effect in the odd nuclei, as illustrated in Ref. \cite{PhysRevLett.81.3599}, whereas $\Delta^ {(3)}(N)$ is impacted by single particle states (see e.g., recent discussion in Ref. \cite{Kreim14} and references therein). The challenging problems thus arise include: One lacks a measure of the pairing gap in the even system, which is mixed with the deformation effect, and the empirical pairing gap for the odd system is related to pairing only in an indirect manner (i.e., the energy loss due to the absence of pairing). The physics becomes even more obscure when abrupt changes occur, e.g., around shell closures.
In practice, as a compromise, the value of $\Delta^ {(3)}(N-1)$ has often been compared to the theoretical pairing gap calculated for the even systems \cite{Lesinski2009,Hil2002,Agb14} and to the OES derived from theoretical binding energies \cite{Agb14,PhysRevC.85.014321,PhysRevC.79.034306}. The direct comparison between the theoretical pairing gap and empirical OES is convenient from a computational point of view since only one single calculation is required, which avoids the complicated handling of the blocking effect in the odd nuclei.

It is known that the theoretical pairing gap, e.g., that from the BCS theory, is not an observable and can not be compared with the empirical OES in a straightforward way. However, they can be quantitatively quite close to each other in most cases and both of them are still important quantities that deserves further attention. 
In particular, within the BCS approach, the corresponding pairing gap is given by
\begin{equation}\label{duv}
\Delta=G\sum_i u_iv_i,
\end{equation}
where $G$ is the pairing strength, and $u_i$, $v_i$ are the standard occupation numbers. This implies that the pairing gaps can serve as a signature of the
change in  two-particle spatial correlation/clusterization, since they are also 
proportional to $\sum_k u_kv_k$. This feature is also responsible for the clustering 
of the four nucleons that eventually constitute the $\alpha$-particle at the 
nuclear surface of heavy nuclei \cite{PhysRevLett.110.242502,PhysRevC.81.064319}.

In this paper we would like to argue that, among the expressions for the OES
studied here, the simple three-point formula $\Delta^{(3)}_{C}(N)$ removes to a large extent the contribution from the varying part of the nuclear 
mean field as well as contributions from other shell structure details and can serve as a reliable indication for the pairing effect in even-$N$ systems. In
other words, $\Delta^{(3)}_{C}(N)$ can be compared to the theoretical pairing gap in a semi-quantitive manner and thus contains fruitful information on the pairing effects. In particular, the abrupt changes in $\Delta^{(3)}_{C}(N)$ do have physical meaning. Moreover, by using $\Delta^{(3)}_{C}(N)$ one can make it more convenient to extract the neutron-proton interaction from binding energy differences \cite{Qi2012436}. We will also show that 
$\Delta^{(3)}_{C}(N)$ is free from the 
Wigner effect for nuclei around $N=Z$ and that its weakening 
in some neutron-rich nuclei may indicate that the di-neutron spatial
correlation itself is weak in those nuclei. 

There are other formulas available for the pairing gap including the so-called four-point and the
five-point formulas. The four-point formula is defined as~\cite{bohr1998nuclear}
\begin{multline}
\label{eq:4point}
\Delta^ {(4)}(N)= \frac{1}{4} [-B(N+1,Z)+3B(N,Z)\\
-3B(N-1,Z)+B(N-2,Z)]\\
=\frac{1}{2}[\Delta^{(3)}(N)+\Delta^{(3)}_C(N)].
\end{multline} 
That is, it measures the average value of $\Delta^ {(3)}$ in adjacent even and odd systems.
The five-point formula is given by~\cite{NuclPhysA.476.1,NuclPhysA.536.20,PhysRevC.65.014311}
\begin{multline}
\label{eq:5point}
\Delta^ {(5)}(N)= \frac{1}{8} [B(N+2,Z)-4B(N+1,Z)\\
+6B(N,Z)-4B(N-1,Z)+B(N-2,Z)]\\
=\frac{1}{4}[\Delta^{(3)}_C(N+2)+2\Delta^{(3)}(N)+\Delta^{(3)}_C(N)].
\end{multline}
The five-point formula is also used in Refs.
\cite{PhysRevC.60.051301,PhysRevC.87.064308,Bender2000}. 
In Refs. \cite{PhysRevC.88.034314,Dobaczewski01032002}, the experimental pairing gap is taken as the average of adjacent ones deduced through the three-point formula as
\begin{eqnarray}
\label{eq:averr}
\begin{aligned}
\Delta_{ave}^{(3)}(N)=\frac{1}{2}\left[\Delta^{(3)}_{C}(N) + \Delta^{(3)}_{C}(N+2)\right],
\end{aligned}
\end{eqnarray}
which is actually also a five-point formula involving the same group of nuclei as $\Delta^ {(5)}(N)$ but with different weights for each nucleus. 
Our calculations show that there is no significant 
difference between the results derived from $\Delta_{ave}^{(3)}(N)$ and 
$\Delta^{(3)}_{C}(N)$ for open-shell nuclei where the pairing gap is a smooth 
function of $N$. For the same reason, $\Delta^{(4)}(N)$ and $\Delta^{(5)}(N)$ show quite a similar behavior for most nuclei. Noticeable differences between $\Delta^{(3)}_{ave}(N)$ and $\Delta^{(3)}_{C}(N)$ 
may be seen where abrupt changes in pairing correlations are expected to 
happen, e.g., around shell closures, which is smoothed out in the former case. 

\begin{figure} [htdp]
\begin{center}
\includegraphics[width=0.45\textwidth]{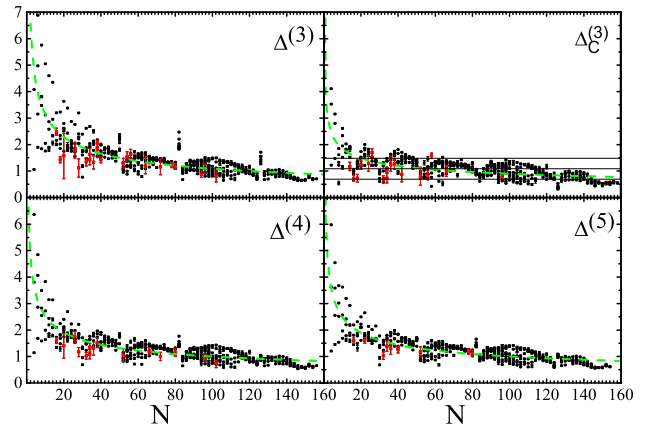}
\end{center}
\caption{\label{fig:4Gaps_errors} Empirical neutron pairing gaps (in MeV) from different OES formulas  for all even-even nuclei as a 
function of neutron number $N$. Results which errors larger than $100$ keV are 
marked in red color. The dashed curves are determined by fitting to the data. 
The solid horizontal lines show the average value of $\Delta^ {(3)}_{C}(N)$ (1.08$\pm$0.40 MeV)
and the corresponding $1\sigma$ uncertainty).
}
\end{figure}

\begin{table*}

\caption{\label{tab:fitt}The fittings of empirical neutron pairing gaps in even-even and even-$N$ odd-$Z$ nuclei from different OES formulas as functions of $A$ and $N$ and the corresponding root-mean-square deviation, coefficients (with 95\% confidence bounds)}
\begin{tabular}{c c c c c}
\hline
\hline
Formula &  even-even &  even-even & even-odd &  even-odd\\
\hline
\multirow{2}{*}{$\Delta^ {(3)}$} & $ (13.43 \pm 1.38) A^{ -0.48 \pm 0.03}$ & $(10.62 \pm 0.8) N^{-0.50 \pm 0.04}$ & $ (7.58 \pm 0.89) A^{-0.41\pm 0.03}$& $ (6.73\pm 0.57) N^{-0.44 \pm 0.02}$\\
& $\sigma= 0.52$  &$\sigma=0.47$& $\sigma=0.46$&$\sigma=0.42$  \\
\hline

\multirow{2}{*}{$\Delta^{(3)}_{C}$} & $(4.03\pm 0.46) A^{-0.28\pm 0.03}$ & $(3.66 \pm 0.33) N^{-0.30 \pm 0.02}$ & $(1.40 \pm 0.27) A^{-0.13 \pm 0.04}$& $ (1.37 \pm 0.23) N^{-0.14 \pm 0.04}$\\
& $\sigma=0.31$ &$\sigma=0.29$ & $\sigma=0.29$&$\sigma=0.28$  \\
\hline

\multirow{2}{*}{$\Delta^ {(4)}$} &  $ (7.37 \pm 0.75) A^{-0.38 \pm 0.02}$ & $  (6.25 \pm 0.49) N^{-0.39 \pm 0.02}$ & $ (3.46 \pm 0.47) A^{-0.28 \pm 0.03}$& $(3.23 \pm 0.35) N^{-0.30 \pm 0.06}$\\
& $\sigma=0.35 $ & $\sigma=0.32$ & $\sigma=0.29$  & $\sigma=0.28$\\
\hline

\multirow{2}{*}{$\Delta^ {(5)}$} & $ (6.93 \pm 0.70) A^{-0.37 \pm 0.02}$ & $ (5.90 \pm 0.45) N^{-0.38  \pm 0.04}$&$(3.94 \pm 0.50) A^{-0.31 \pm 0.03}$& $(3.46 \pm 0.35) N^{-0.32 \pm 0.05}$\\
& $\sigma=0.34$ &$\sigma=0.31$   & $\sigma=0.26$ & $\sigma=0.25$  \\
\hline
\hline
\end{tabular}
\end{table*}

The empirical pairing gaps obtained by various OES formulas are shown in Fig. 
\ref{fig:4Gaps_errors} for which the nuclear binding energies are extracted 
from Refs. \cite{ChinPhysC.36.1157, Nature.498.346349}. The results are fitted by the expression $aN^{\alpha}$ where $a$ and $\alpha$ are parameters to be determined. Our calculations show that this expression performs equally well as the usual $A$-dependence fit, as can be seen in Table \ref{tab:fitt} where
the results of fitting a power function to the different mass formulas as a function of neutron number, $N$, and mass numbers, $A$, are given.
It is also seen that the dependence of the gap
upon the neutron or mass number is weakest for $\Delta^ {(3)}_C(N)$. This is consistent with the 
conclusions of Refs. \cite{Friedman2009,Hil2002} that the pairing gap should not show significant mass dependence.
 Only in nine cases one has $\Delta^ {(3)}_C(N)$ larger than 2 MeV. They correspond to 
$^8$Be (4.11), $^{10}$Be (2.57), $^{10}$C (3.53), $^{12}$C (2.8), 
$^{14}$O (3.15), $^{20}$Ne (2.6), $^{22}$Mg (2.3), $^{26}$Si (2.02) and 
$^{44}$Ti (2.0) (within parenthesis are the corresponding pairing gaps in MeV).

The  mean values of the pairing gaps corresponding to the different
expressions given above are (in MeV) $\Delta^ {(3)}(N)$ = 1.46, 
$\Delta^ {(3)}_C(N)$ = 1.08, $\Delta^ {(4)}(N)$ = 1.26 and 
$\Delta^ {(5)}(N)$ = 1.26. We also evaluated the empirical pairing gaps for the proton for which one obtains $\Delta^ {(3)}_C(Z)=1.01\pm0.36$ MeV, which contains a smooth contribution from the Coulomb field. 

In Table  \ref{tab:fitt} we also included calculations for the even-$N$ odd-$Z$ nuclei. However, it should be mentioned that the empirical OES $\Delta^ {(3)}_C(N)$ thus extracted contains a sizable negative contribution from the residual neutron-proton interaction between the odd particles in the intermediate odd-odd nuclei.

As indicated in Eqs. (1), (3) and (4), all these OES formulas contain constantly a contribution from the mean field which peaks at the shell closure and persists in open-shell nuclei \cite{PhysRevLett.81.3599}. For the same reason, the OES formula Eq. (2) may not be applicable to evaluate the pairing gaps in odd-$N$ nuclei.

\begin{figure}[htdp]
\begin{center}
\includegraphics[width=0.45\textwidth]{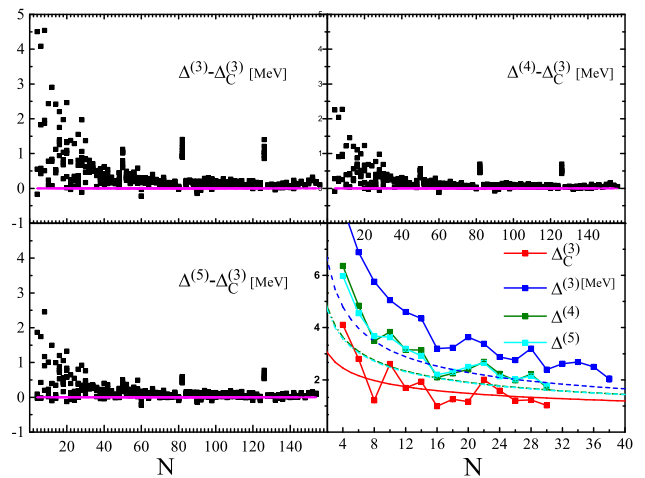}
\end{center}
\caption{\label{fig:Diff_Gaps_subplot3} Differences between the neutron gaps derived from different gap formulas in Fig. 1 with respect to $\Delta^{(3)}_{C}(N)$. The right bottom figure shows the neutron gaps for $Z=N$ nuclei in comparison with their corresponding fitted curves from Fig. 1. }
\end{figure}

\begin{figure}[htdp]
\begin{center}
\includegraphics[width=0.45\textwidth]{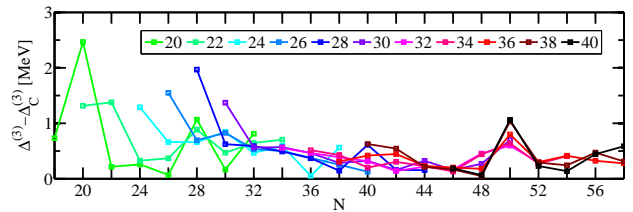}
\end{center}
\caption{\label{fig:Diff_Gaps_subplot4} Same as Fig. \ref{fig:Diff_Gaps_subplot3} but for the difference between $\Delta^{(3)}(N)$ and $\Delta^{(3)}_{C}(N)$ in selected isotopic chains. }
\end{figure}

The differences between the various gap formulas and $\Delta^{(3)}_{C}$ are 
plotted in Figs. \ref{fig:Diff_Gaps_subplot3} and \ref{fig:Diff_Gaps_subplot4}. The dispersal of the data below 
$N\sim30$ ($A\sim50$) as well as around shell closures can be an indication of the significant mean-field 
contribution to the gaps in above regions, which is expected to show a $A^{-1}$ dependence \cite{Friedman2009}. In this context it is worthwhile to point out that the differences between 
$\Delta^ {(3)}(N)$ and $\Delta^ {(3)}_{C}(N)$ was found to be a consequence
of the gap between the single-particle energies of the 
corresponding neighboring orbitals \cite{PhysRevLett.81.3599}.  As can be seen in Fig. \ref{fig:Diff_Gaps_subplot4}, the differences between 
$\Delta^{(3)}_{C}$ and $\Delta^{(3)}$ show quite small values for nuclei that can be reasonably described within a single-$j$ shell or nearly-degenerate systems, e.g., for nuclei with $N$ between 20 and 28 ($f_{7/2}$) as well as $N$ below ($g_{9/2}$) and above ($d_{5/2},g_{7/2}$) $N=50$. For these systems, as we will show below, the difference between $\Delta^{(3)}_{C}$ and $\Delta^{(3)}$ is mainly induced by the Pauli and particle blocking effects.

$\Delta^{(3)} (N)$ show quite large values around $N=Z$ 
nuclei with $Z=4-14$ and 22. One may suspect that $\Delta^{(3)} (N)$ is still 
contaminated by the Wigner effect, which refers to the additional binding 
gained in $N=Z$ nuclei. This is indeed the case for all the other three OES 
formulas where, as seen in the lower right panel of Fig. 2, the calculated gap 
values for $N=Z$ nuclei are systematically much larger than the average values. 
On the other hand, the calculated $\Delta^{(3)}_{C} (N)$ values for $N=Z$ 
nuclei follow nicely the average behavior, as it was also shown in Ref. 
\cite{Qi2012436}, even though relatively large fluctuations are present.
These fluctuation may be due to dramatic changes in the mean field when 
going from 
the daughter ($N-2$) system to the mother nucleus. For instance, they may be due to
quite different deformation properties. An illustration of this is provided
by the nucleus $^8$Be, which shows a deformed two-$\alpha$-particle cluster 
structure while $^6$Be (as well 
as the mirror nucleus $^6$He) is spherical. This is also the case for the 
nuclei $^{20}$Ne and $^{44}$Ti which are expected to gain additional binding 
due to the strong neutron-proton quadrupole correlation.

To illustrate the origin of the OES and the difference between different formulas, we start with the simple seniority model. For a system with $n$ identical particles in a single $j$-shell, the binding energy can be solved analytically in the seniority scheme. The Hamiltonian for such a model can be written as \cite{ring2004nuclear},
\begin{eqnarray}
\begin{aligned}
\label{eq:seniority}
H =-G\sum_{m,m^{\prime}> 0}a^{\dagger}_{m}a^{\dagger}_{-m}a_{-m^{\prime}}a_{m^{\prime}}=-GS_{+}S_{-}
\end{aligned}
\end{eqnarray}
where $S$ represents the quasi-spin operator, and $a^{\dagger}_{m}$ and $a_{m}$ are creation and annihilation operators. $G$ is the strength of the pairing interaction. The energy of the state with seniority $v$ can be written as
\begin{eqnarray}
\label{eq:energySin2}
E(n)&=&-G\frac{n-v}{4}(2j+3-n-v)\\
&=&\frac{n(n-1)}{4}G-\frac{v(v-1)}{4}G-\frac{1}{2}(n-v)(j+1)G\nonumber
\end{eqnarray}
If one assumes  $v=0$ for the ground state of even-even system and $v=1$ for that of the odd system, the expression above can be simplified as
\begin{eqnarray}
\label{eqs}
E(n)&=&\frac{n(n-1)}{4}G-\left[\frac{n}{2}\right](j+1)G,\\
&=& \left[\frac{n}{2}\right]\left(\left[\frac{n}{2}\right]-1\right)G +\delta_{v,1}\left[\frac{n}{2}\right] G+\left[\frac{n}{2}\right]E_2\nonumber
\end{eqnarray}
where $[n/2]$ denotes the largest integer not exceeding n/2 and corresponds to the total number of $v=0$ pairs. 
The nonlinear term in the equation above, which is proportional to $n(n-1)$ or more exactly $\left[\frac{n}{2}\right]\left(\left[\frac{n}{2}\right]-1\right)$, is related to the energy loss due to the Pauli effect. The $\delta_{v,1}$ term in above equation indicates the energy loss in the $v=1$ odd system due to the the particle (Pauli) blocking effect: The unpaired particle blocks the scattering of other pairs to its own level. In practice, the odd system corresponds to a even system with one less particle and with a lower degeneracy.

\begin{figure}[htdp]
\begin{center}
\includegraphics[width=0.45\textwidth]{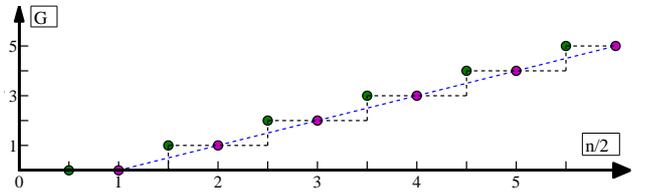}
\end{center}
\caption{\label{sch}A schematic plot for the competition between the Pauli blocking effect and pairing blocking effect of the unpaired odd particle. }
\end{figure}

 $E_2$ is the energy of one pair.
This term is the one that contributes to the theoretical OES and for systems with even $n$ we have 
\begin{equation}
\Delta^ {(3)}_{C}(n)=-\frac{1}{2}E_2 = \frac{1}{2}(j+\frac{1}{2})G.
\end{equation}
It should be emphasized that a striking feature one thus finds is that the two blocking terms cancel each other in $\Delta^ {(3)}_{C}(n)$ and they do not contribute to the pairing gap. A schematic picture is given in Fig. \ref{sch} to illustrate this point.
However, as implied from the figure, this is not the case for $\Delta^ {(3)}(n)$ for which one has,
\begin{equation}
\Delta^ {(3)}(n)=-\frac{1}{2}E_2 +\frac{1}{2}G,
\end{equation}
where the term $G/2$ is due to the superfluous contribution from the blocking effects. This unwanted term is the main origin of the differences between $\Delta^ {(3)}_{C} $ and $\Delta^ {(3)}$, as shown in Fig. \ref{fig:Diff_Gaps_subplot4}, in nuclei that can be well described within a single $j$ shell.

For non-degenerate systems the pairing collectivity
manifests itself through the correlated contribution from many configurations, which is induced by the non-diagonal matrix elements of the pairing interaction in a shell-model context.
For two particles in a non-degenerate system with a constant pairing, the energy can be evaluated through 
the well known relation,
\begin{equation}
\label{dren}
    G\sum_{i}\frac{2j_i+1}{2\varepsilon_i-E_2}=2.
\end{equation}
The corresponding wave function amplitudes are given by 
\begin{equation}
X_i=N_n\frac{2j+1}{2\varepsilon_i-E_2}
\end{equation}
where
$N_n$ is the normalization constant. All amplitudes $X_i$ contribute to the 
two-particle clustering with the same phase due to the strongly attractive nature 
of the pairing interaction. The correlation energy induced by 
the monopole pairing corresponds to the difference
\begin{equation}
\Delta^ {(3)}_{C}=\varepsilon_{\delta}-\frac{1}{2} E_2,
\end{equation}
where $\delta$ denotes the lowest orbital. As the gap $\Delta$ increases the 
amplitude $X_i$ becomes more dispersed, resulting in stronger two-particle spatial
correlation. This difference, or more exactly 
$\Delta-1/2G$ with the self energy removed, is an important measure of the 
two-particle spatial correlation at the surface, reflected in a corresponding  clustering of the
two nucleons forming the pair (see, e.g., Ref. \cite{PhysRevC.81.064319}).
This clustering induces an increase in the strength of the corresponding 
pair-transfer reaction. 

\begin{figure}[htdp]
\begin{center}
\includegraphics[width=0.4\textwidth]{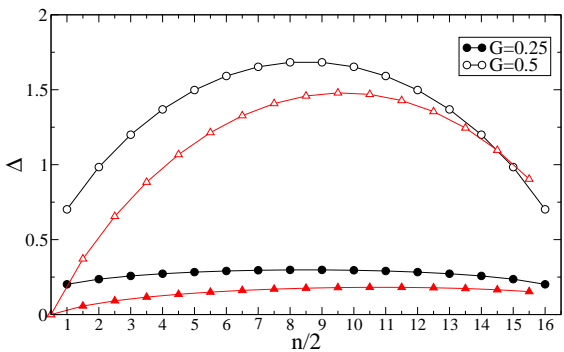}
\end{center}
\caption{\label{fig3} Pairing gaps $\Delta^ {(3)}_{C}$ for a equally spaced douby degenerate system with 16 levels and $G=0.25$ (solid circle) and 0.5 (open circle). The triangles denote the contribution from the particle blocking effect, $\delta$.}
\end{figure}

As an example, in Fig. \ref{fig3} we consider a simple model with a set of equally spaced levels with double degeneracy. We consider 16 levels and the single-particle energy is taken as $\varepsilon_i=i$. The total energy is obtained by diagonalizing the corresponding Hamiltonian matrix. It is found that the total energy for such a system follows closely a relation similar to Eq. (\ref{eqs})
\begin{eqnarray}
E(n)
&\simeq& \left[\frac{n}{2}\right]\left(\left[\frac{n}{2}\right]-1\right)\mathcal{G} +\delta_{v,1}(\varepsilon_b+\delta)\nonumber\\
&&+\left[\frac{n}{2}\right]E_2,
\end{eqnarray}
where $\varepsilon_b$ is the single-particle energy of the unpaired orbital for a odd system, $\delta$ is the energy loss due to the related blocking effect. It contributes significantly to the OES and determines its evolutions as a function of pair number. $\mathcal{G}$ is a parameter that is related to the pairing strength and level density. One has $\mathcal{G}\simeq1.01$ and 1.06 for $G=0.25$ and 0.5, respectively.

Within the simple BCS context, the separation energy is approximately given by $S_{2n} (N)\approx -2\lambda$ and 
$S_n(N-1)\approx -\lambda-\Delta$, where $\lambda$ is the Fermi energy. One
thus gets
\begin{equation}
\Delta^{(3)}_C (N)\approx\Delta.
\end{equation}
But one should bear in mind, as the exact solution of the pairing Hamiltonian in Fig. \ref{fig3} indicates, that there is still a sizable difference between the blocking effect and the OES especially for small systems with only few particles.

From a macroscopic point of view, there may be a residual 
contribution to $\Delta^{(3)}_{C}$ from the symmetry energy (expected to be 
negative) and other non-linear terms of the binding energy \cite{PhysRevC.88.064329}. Our calculations with the liquid drop model show that the average residual contributions are around -60 keV and -200 keV for neutron and proton pairing gaps, respectively. The later case is dominated by a smooth contribution from the Coulomb field.
The main argument for the usage of $\Delta^{(5)}$ instead is that the smooth non-linear terms in the binding energy can be canceled up to the fourth order \cite{bohr1998nuclear,PhysRevC.65.014311,Bender2000}. That is, it removes the liquid-drop contribution to the OES to the largest extent. However, as mentioned in Ref. \cite{PhysRevC.88.064329}, the disadvantage is that the odd-even effects may be diminished as a result of the averaging over nuclei further apart. 

It should also be mentioned that there is an additional contribution to the binding energy  from the diagonal matrix element with $m=m'$ (the so-called self-energy) which is equal to $-\left[n/2\right]G$. In the ideal case, one should have this contribution removed and the pairing gap for a single-$j$ system will be of the form
$\Delta =-E_2 - G/2 = 1/2(j-1/2)G. 
$ For heavy nuclei the BCS coupling constant roughly takes $27/A$ which gives a  contribution to $\Delta^ {(3)}_{C}$ that is comparable but opposite to that of the smooth contribution from the nonlinear terms mentioned above. In the other words, one may expect that these two unwanted contributions largely cancel each other. As a result, if one assume that both contributions to the binding energy have been taken into account by the HF configuration,  for the neutron pairing gap one is expected to have \cite{PhysRevLett.81.3599}
\begin{equation}
\Delta^{(3)}_C (HF)=0.
\end{equation}

All the variety of OES formulas studied above have their advantages and disadvantages (see, earlier discussions in Refs. \cite{PhysRevLett.81.3599,PhysRevC.65.014311,Bender2000}). We will focus on the simple $\Delta^{(3)}_{C}$ particular for two reasons: It is the one that may  be least contaminated  by the nuclear shell effect and it involves only three nuclei. These are important for our study of unstable nuclei where the experimental data are scarce wheras the unknown shell evolution can play a decisive role. For open shell nuclei, the uncertainty as related to the different definiation of the OES is much smaller as compared to the uncertainty induced by our limitted understanding of the density functional and the shell structure of dripline nuclei.
Our aims are to understand the pairing properties of neutron dripline nuclei 
and to extract reliable information from binding energies based on the 
gap $\Delta^{(3)}_C (N)$. We hope that this may shed light on our 
understanding of the stability mechanisms as well as probing
the importance of di-neutron spatial correlations in these nuclei. It is theoretically 
suggested that di-neutron correlations may be 
enhanced in some situations such as in a low density region of nuclear matter 
and in the surface of finite neutron-rich nuclei \cite{PhysRevC.88.034314,PhysRevC.88.054308,Pil07,Mat05}. 
To explore this point further we have done systematic calculations 
in  semi-magic neutron-rich (dripline) nuclei by using the 
Hartree-Fock-Bogoliubov (HFB) formalism, which is expected to be more 
reliable than the simple BCS approach.
In the standard HFB formalism, the Hamiltonian is reduced to
the mean field in the particle-hole channel and the pairing field in 
the particle-particle channel. The HFB equations are,
\begin{equation}
\label{eq:H}
 \begin{pmatrix}
 (H-\lambda) & \Delta \\
 -\Delta^{*} & -(H-\lambda)^{*}
 \end{pmatrix}\begin{pmatrix}
	 U_{k}\\
	 V_{k} 	\end{pmatrix}= E_{k} 
	\begin{pmatrix}
	 U_{k}\\
	 V_{k}
	\end{pmatrix} , 
\end{equation}
where $U_{k}$ and $V_{k}$ are the two components of the single quasi-particle 
wave functions. We evaluated the 
coordinate-space solutions by using the HFB solver HFBRAD in a spherical box 
\cite{Bennaceur200596}. In the particle-hole channel we used the Skyrme 
functional with the SLy4 parameter set \cite{Chabanat1998}. In the
particle-particle channel we used the zero-range $\delta$ interaction given
by \cite{Bennaceur200596}
\begin{equation}
\label{eq:surface}
V_{pair}(\textbf{r},\textbf{r}^{\prime})= V_{0}\left(1-\eta\frac{\rho(\textbf{r})}{\rho_{0}}\right) \delta (\textbf{r}-\textbf{r}^{\prime})
\end{equation}
Here $V_{0}$ is the pairing strength, $\rho(\textbf{r})$ is the isoscalar 
nucleonic density and $\rho_{0}=0.16 fm^{-3}$. $\eta=0$ and 1 correspond to the 
volume and surface pairings, respectively. We have done calculations by using 
different pairing interactions but for simplicity only results corresponding
to the mixed pairing with $\eta=0.5$ are shown. The pairing strength is fitted to give a 
mean neutron gap of $1.31$MeV in ${}^{120}$Sn and no $A$ or isospin dependence 
is considered. Calculations are done in a box with the size $r=30$ fm. Only quasiparticle states with energy lower than 60 MeV are taken into account.

\begin{figure}[htdp]
\includegraphics[width=0.4\textwidth]{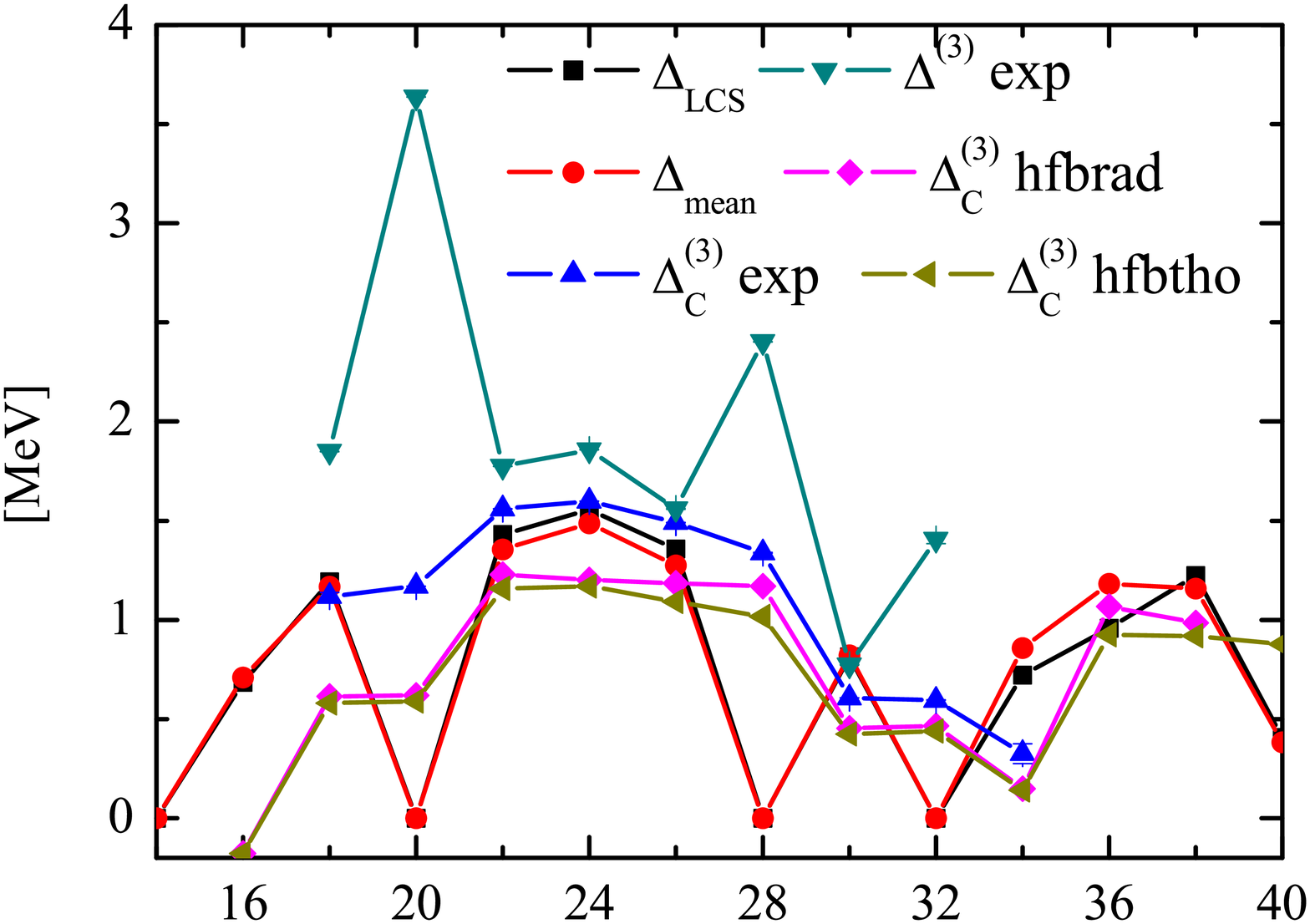}
\includegraphics[width=0.38\textwidth]{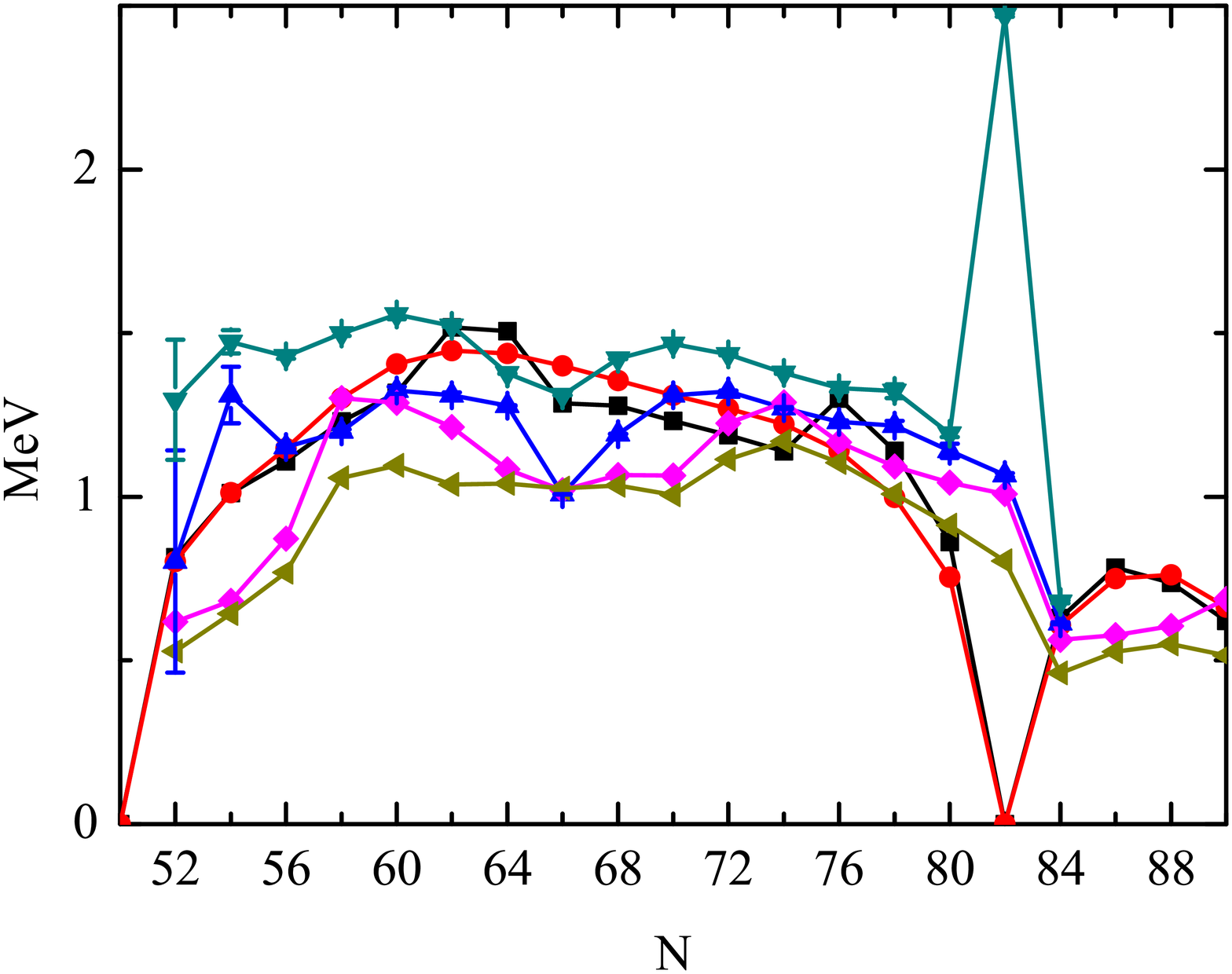}
\caption{\label{fig:TinCalcium} Calculated $\Delta_{LCS}$ and $\Delta_{mean}$ for Ca (upper) and Sn (lower) together with the experimental $\Delta^{(3)}_{C}$ and calculated $\Delta^{(3)}_{C}$ by using the HFRABD and HFBTHO codes. (explained in the text).}
\end{figure}

Besides OES from calculated binding energies, two different theoretical gaps will be compared with the experimental pairing 
gap $\Delta^{(3)}_C$: The canonical gap $\Delta_{LCS}$, which consists of
the diagonal elements 
of the pairing-field matrix for the lowest canonical state (LCS), and the 
average gap $\Delta_{mean}$, that is the average value of the pairing fields 
\cite{Bennaceur200596}.

As typical examples, in Fig. \ref{fig:TinCalcium} we plotted $\Delta_{LCS}$ and $\Delta_{mean}$ calculated for calcium and tin isotopes, which have been intensively studied quite recently from different perspectives. The results are compared with experimental $\Delta^{(3)}_{C}$ and calculated $\Delta^{(3)}_{C}$  which is the pairing gap calculated from the HFB binding energies by using Eq. (2). In the latter case, the binding energies are calculated using the code HFBRAD (in the coordinate representation
within the spherical symmetry) \cite{Bennaceur200596} and compared with those from the code HFBTHO (in the axial deformed harmonic oscillator basis) \cite{Stoitsov20131592} as a cross-check for our calculations. The two calculations should give very similar results if the influence of both the continuum and deformation is negligible. Indeed, as can be seen from the figure, the results from the two codes are very close to each other in most cases and come even closer if spherical symmetry constraint is applied in the latter case. In both codes calculations for odd-$N$ isotopes are done within the blocking approximation using the equal
filling approximation \cite{Per2008,Sch2010}. In practice, we have done calculations by blocking all possible quasiparticle orbitals around the Fermi surface and the one that gives the highest binding energy is chosen. A self-consistent treatment of the blocking for the one-quasiparticle HFB state is done in Ref. \cite{Ball14} by taking into account beyond mean field effects using the  Generator Coordinate Method.
In general the values of calculated $\Delta^{(3)}_{C}$ are closer to the experimental $\Delta^{(3)}_{C}$ 
than the two theoretical HFB pairing gaps, which vanish at closed shell.

The results for $\Delta_{LCS}$ and $\Delta_{mean}$ are practically the same in 
most cases in Fig.  \ref{fig:TinCalcium}. But it should be mentioned that both definitions of the theoretical pairing gap have their advantages and disadvantages. 
The use of $\Delta_{LCS}$ as a measure of the pairing gap is questioned in Ref. \cite{PhysRevC.65.014311} and in Refs. \cite{PhysRevC.88.034314,Agb14} for dripline nuclei.
In the relativistic Hartree-Bogoliubov calculations for selected isotopic and isotonic chains with a finite-range pairing force as presented in Ref. \cite{Agb14}, it is seen that $\Delta_{mean}$ are systematically smaller than $\Delta_{LCS}$ and come closer to the $\Delta^ {(3)}$ and $\Delta^ {(5)}$ indicators. In Ref. \cite{PhysRevC.88.034314}, it is also shown that the neutron $\Delta_{LCS}$ vanishes at the drip line for nearly all semi-magic isotopic chains studied whereas the gap $\Delta_{mean}$ can persist in some cases.
On the other hand, $\Delta_{mean}$ and $\Delta_{LCS}$ can give quite similar prediction for bound nuclei. Calculations in Ref. \cite{PhysRevC.88.034314} show that $\Delta_{LCS}$ can be closer to the experimental average gap Eq. (\ref{eq:averr}) than $\Delta_{mean}$. In the relativistic mean field calculations \cite{Agb14}, the calculated mean gaps and OES agree pretty well with each other. 
In Ref. \cite{Lit2005}, the mean gap and OES from calculated binding energies are compared with the empirical OES for even-even hafnium isotopes, where it is found that quantatively the mean gap reproduces better the trend in experimental data.

The level spacing is crucial for pairing correlation which strongly depends on the shell structure \cite{Afa15} and on the deformation \cite{PhysRevLett.81.3599}.
A general calculations  including the deformation effects are done recently in Ref. \cite{Afa15} with a  separable pairing interaction of finite range. It is shown that the differences in the underlying single-particle
structure
represent the major source of uncertainty
in the prediction  of drip line and the pairing properties in neutron rich nuclei
depend substantially on the different covariant energy density functionals. In particular, the emergence of deformation driving intruder orbitals can lower the chemical potential and make the system bound whereas  extruder orbitals with high $\Omega$ values can make the system unbound. A systematic calculation on the possible effect of the density dependence of the zero-range pairing force on the pairing correlation of dripline nuclei is also shown in Ref. \cite{Cha15}.

\begin{figure}[htdp]
\includegraphics[width=0.4\textwidth]{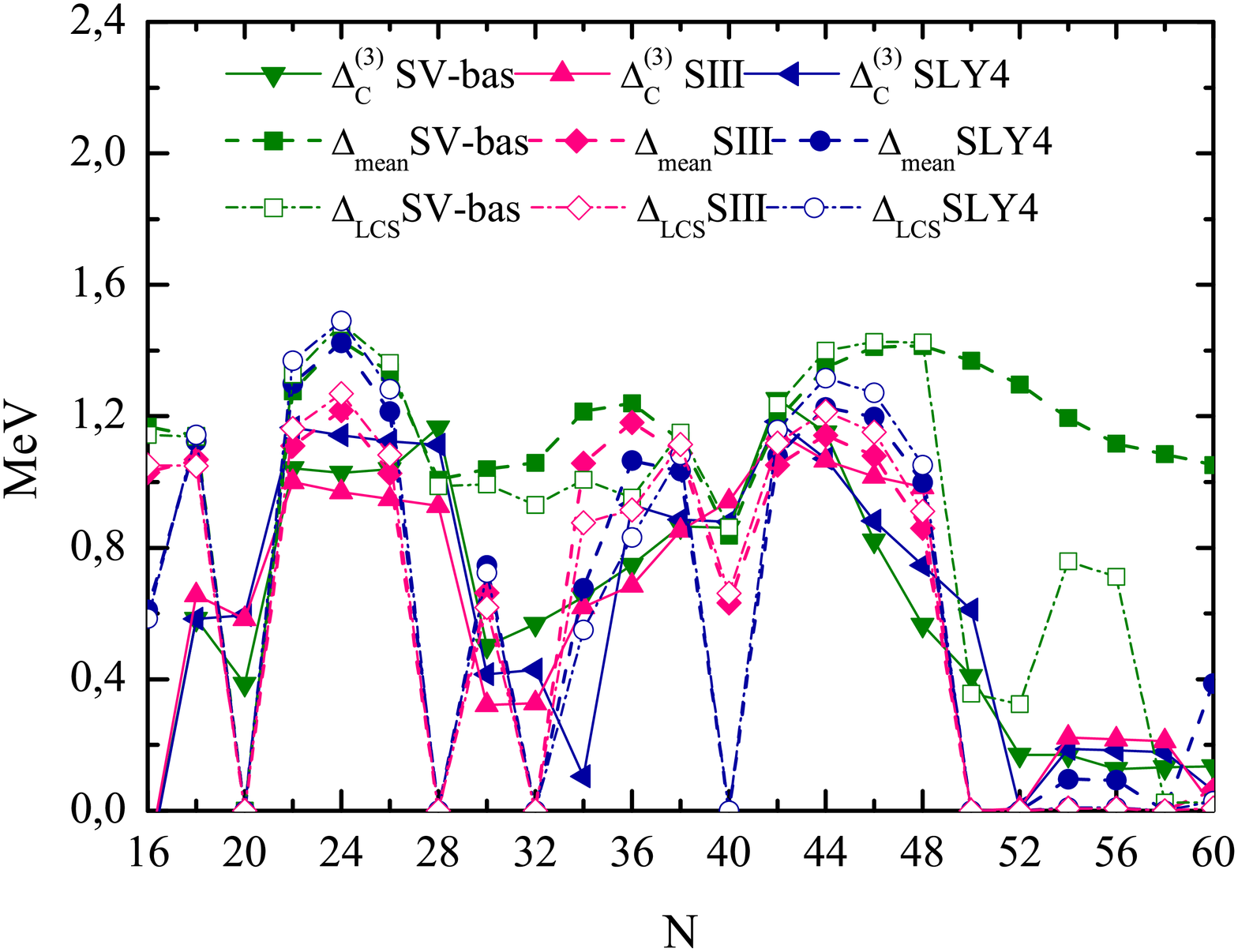}
\includegraphics[width=0.4\textwidth]{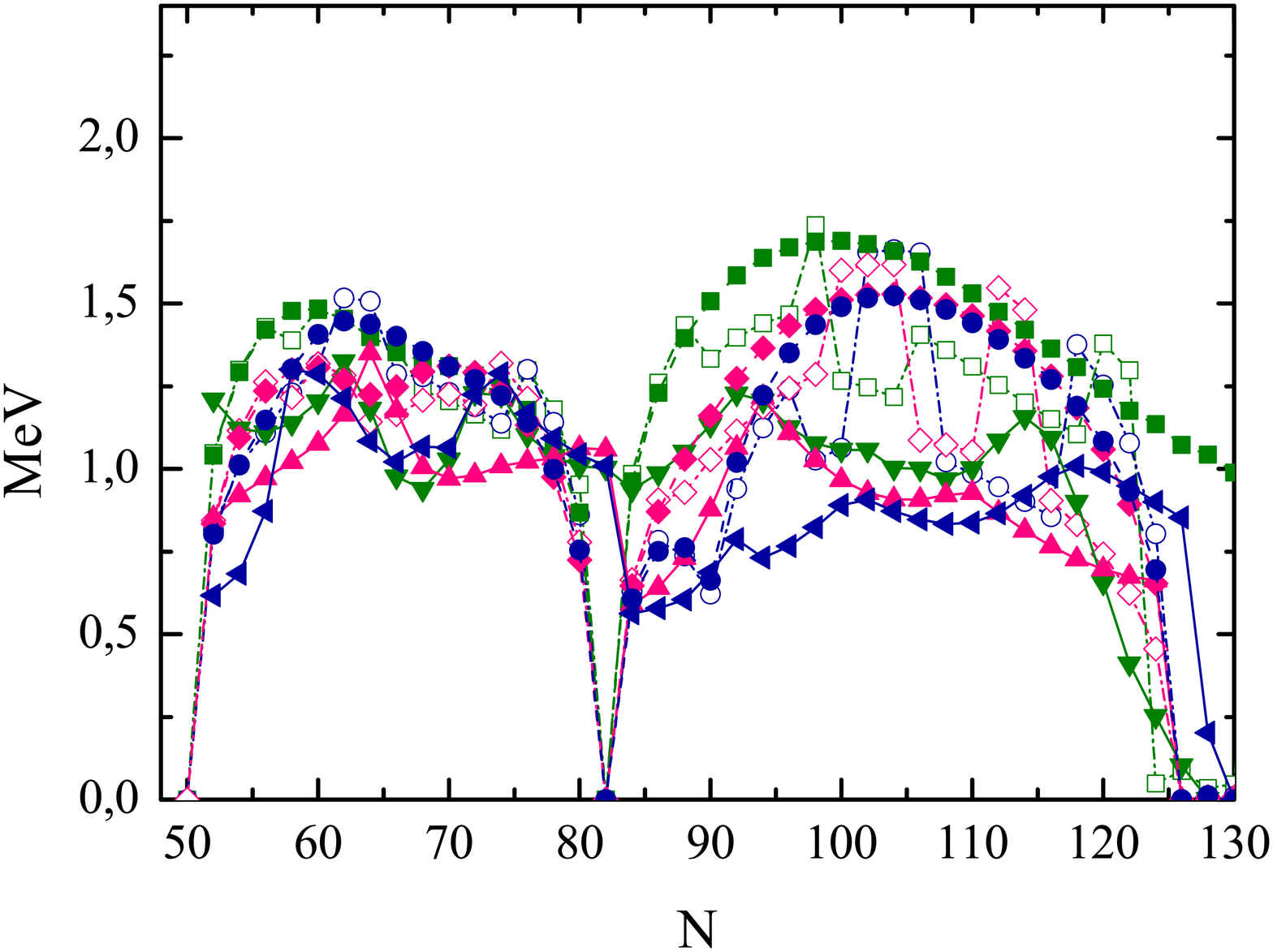}
\caption{\label{fig:TinCalcium2} Calculated $\Delta_{LCS}$ and $\Delta_{mean}$ for Ca (upper) and Sn (lower) isotopes upto the neutron dripline with three different Skyrme parameter sets in comparison with the corresponding calculated OES $\Delta^{(3)}_{C}(N)$.}
\end{figure}

To explore the dependence of our calculations on the functionals, 
In Fig. \ref{fig:TinCalcium2} we extended our calculations for Ca and Sn isotopes shown in Fig. \ref{fig:TinCalcium} to the neutron dripline.
Calculations are done with the SIII \cite{siii}, SLy4 \cite{Chabanat1998} as well as the recently proposed SV-bas \cite{svb} Skyrme functionals. The two-neutron dripline occurs around $N=48$ and $N=126$ for those two isotopic chains. It is thus found that the calculated OES agrees well with the two pairing gaps for known nuclei but noticable differences are seen as the neutorn number increases. In particular, big differences among different calculations are seen for Ca isotopes around $N=32$ in relation to the different predictions of the emergence of the $N=32$, 34 and 40 subshells. Significant deviations between $\Delta_{mean}$ and $\Delta_{LCS}$ are seen in SV-bas calculations for nuclei beyond the dripline. In fact,
Ref. \cite{PhysRevC.88.034314} has shown that 
there is a distinct difference between the mean gap and the lowest canonical gap 
beyond drip-line for calculations with surface-peaked pairing interaction. A similar deviation may happen in calculations with the mixed pairing interaction, as in the case shown in Fig. \ref{fig:TinCalcium2}.
However, this is usually not the case for volume interactions.

It should be mentioned that the time-odd field is not considered in the HFBTHO and HFBRAD calculations shown in the figure, which may affect the binding energy of the odd-$A$ nucleus and the corresponding OES. The time-odd field can contribute to the ground state mean fields of odd-$A$ and odd-odd nuclei (breaking the Kramer's degeneracy) as well as in nuclear rotation and other dynamic processes \cite{Bender02}. The time-odd field, which is still relatively poorly understood and functional dependent, is neglected in most relativistic and non-relativistic mean field calculations. The possible influence of the time-odd field in light $N = Z$ nuclei is discussed in Ref. \cite{Sat99}. 
Systematic Skyrme Hartree-Fock plus BCS calculations with the time-odd field for odd-$A$ nuclei between$16\leq Z \leq 92$  are presented recently in Ref. \cite{Pot10}.
It was shown that the influence of the time-odd field is generally small and decrease rapidly with increasing mass number. In particular, there is no indication that the deviation between experimental and theoretical OES can be improved by the inclusion of the time-odd field \cite{Pot10}, from which the fluctuation induced is much smaller than that from the different functional. Skyrme HFB calculations with the time-odd field are also done for the whole mass table in Ref. \cite{Mar09}  and for rare earth nuclei with  $63\leq Z\leq 75$ and $78\leq N \leq 104$ in Ref. \cite{Sch2010}. 
It is found that the average energy shift as induced by the time-odd field is only 50 keV but with a large deviation of 42 keV and hence the equal filling approximation is precise for most practical calculations. A similar conclusion may be draw from Gogny HFB calculations where full calculations with the time-odd field give very similar results to those from the filling approximation and the inclusion of the time-odd field hardly improves the gap description \cite{Rob12}. Recent systemtic calculations on the effect of the time-odd field in the relativistic mean field approach can be found in Refs. \cite{Afa10,Afa11,Xu14} and references therein. Ref. \cite{Afa10} found that the time-odd field in the relativisitic mean field approach always induces an additional binding but weakly affects the relative energies of different quasiparticle states
in medium and heavy mass nuclei. The influence of that attractive contribution from
the time-odd mean field to the odd nuclei  into OES can be
of up to 10\% in light nuclei and of around 5-6\% in heavy
nuclei. However, even in that case, the required modification of the strength of pairing force is modest \cite{Afa10}, which is significantly smaller than earlier expectations. In Ref. \cite{PhysRevC.65.014311}, it is suggested that the $\Delta^{(3)}_{C}(N)$ can be a measure of the sole pairing gap if the contribution from the time-odd reversal symmetry breaking field
cancels out the contribution from the smooth part of the total  energy.

\begin{figure}
\begin{center}
\includegraphics[width=0.45\textwidth]{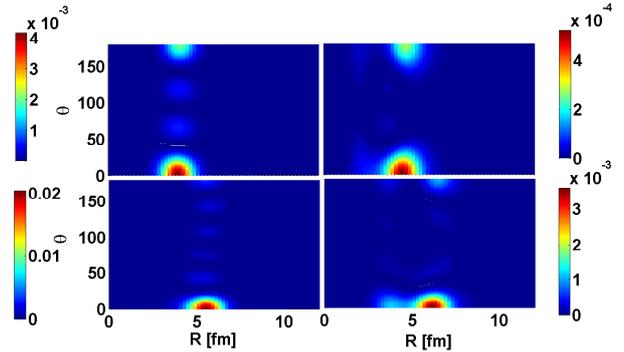}
\end{center}
\caption{\label{fig:twopar_subplot} Upper: The two-neutron spatial correlation plots for $^{46}$Ca (left) and $^{54}$Ca (right).  Lower: Same as upper but for $^{128}$Sn (left, 4 holes) and $^{136}$Sn (right, 4 particles). Calculations are done with the SLy4 force. Notice that the scale is different.}
\end{figure}

In Ref. \cite{PhysRevC.88.034314} it is mentioned that for spherical nuclei where the drip line 
occurs around neutron closed shells the pairing gap is expected to 
be reduced. Our calculations show that this is indeed the case for neutron-rich Ca and Sn isotopes. Our 
understanding of the shell structure in Ca isotopes have been significantly 
extended recently (see, e.g., Refs. \cite{Nature.498.346349,Xu2013247,PhysRevC.89.034316}). 
$N=32$ and 34 have been shown to be new magic numbers in Ca isotopes, a
conclusion which is 
supported by HFB calculations \cite{PhysRevC.89.034316}. The $\Delta^{(3)}_{C}(N)$  values 
for $^{50,52}$Ca are much smaller than those for nuclei in the $0f_{7/2}$
shell. 
The pairing gap for $^{54}$Ca is expected to be even smaller. In the 
upper panel of Fig. \ref{fig:twopar_subplot},  following the same procedure as 
in Ref. \cite{PhysRevC.81.064319}, we evaluated in the canonical basis the two-neutron spatial correlations 
in $^{54}$Ca and compared it with that in $^{46}$Ca. The results are plotted as a 
function of the angle $\theta$ between the two neutrons and the radius
 $R=r_1=r_2$ where $r$ is the radius of the single-particle wave function. For 
strongly correlated wave functions, the two-neutron spatial correlation is expected to 
peak at $\theta=0$ at the surface. As can be seen from the figure, the peak 
in $^{46}$Ca is much stronger than that in $^{56}$Ca.

\begin{figure}
\begin{center}
\includegraphics[width=0.45\textwidth]{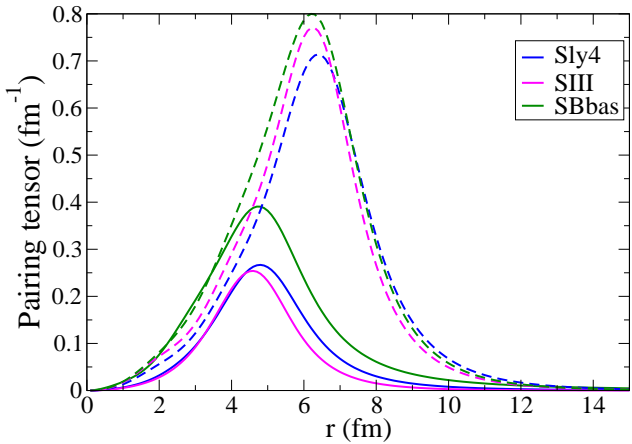}
\end{center}
\caption{\label{fignnn} The pairing tensor as a function of radius $r$ for neutron-rich nuclei $^{78}$Ca (solid line) and $^{160}$Sn (dashed line) as calculated from different Skyrme parameterizations.}
\end{figure}

The spatial correlation of the neutron pair can also be inferred from the pairing/abnormal density $\tilde{\rho} (r)$. in Fig. \ref{fignnn}, we  plotted the calculated pairing tensor which is defined as $\kappa (r) =r^2\tilde{\rho} (r)$ and is related to the pair transfer form factor,  for two nuclei $^{78}$Ca (solid line) and $^{160}$Sn. $^{78}$Ca is at the neutron dripline for which the calculated pairing tensors and pairing gaps from difference functionals can be quite different. The nucleus $^{160}$Sn is within the dripline. The pairing tensors calculated from the three functional are quite close to each other. But as can be seen from Fig. \ref{fig:TinCalcium2}, there is a significant difference between the theoretial mean gap $\Delta_{mean}$ and the OES from the theoretical binding energies. A similar deviation is also seen in Ref. \cite{Bender2000}. The origin for such a large deviation is not exactly known yet. It may be related to the fact that, for Sn isotopes in this region, the chemical potential $\lambda$ is pretty close to zero and the quasiparticle orbitals involved in the blocking calculations are all unbound. For such cases, the theoretical OES and gaps have to be carefully interpreted.

There is a noticeable kink at $N=66$ for tin isotopes which may be related to 
the occupancy of the low degeneracy orbital $s_{1/2}$. 
A sudden drop occurs in the experimental pairing gaps in the Sn isotopes when 
going from $N=82$ to $N=84$. This can be reproduced by the HFB calculations 
with the 
volume and mixed pairing interactions \cite{PhysRevLett.109.032501}. That drop 
is related to the reduced level density above N=82. That is, there
is a noticeable gap between the $1f_{7/2}$ orbital and the higher-
lying ones.  which may result in a $N=90$ sub shell closure \cite{Xu2013247}. 
The reduced pairing collectivity in these nuclei can be clearly seen in the 
two-neutron spatial correlation plots given in the lower part of Fig. 
\ref{fig:twopar_subplot}, where we compared $^{128}$Sn, which 
have four holes in the core $^{132}$Sn, and $^{136}$Sn, with four particles 
above the core.
As can be seen from the figure, the peak in $^{128}$Sn is around one order of magnitude stronger than that in $^{136}$Sn. 

HFB describes very well the pairing gap as well as the two-particle spatial correlation in open-shell nuclei. A known problem is that both the BCS
and the HFB condensates collapse at closed shell, since in this case all 
orbitals are fully occupied 
(c.f., Eq. (13)). In fact, as the systematic calculations in Ref. 
\cite{PhysRevC.79.034306} shows, roughly 20-30\% of the known nuclei contain 
collapsed BCS or HFB condensates. It is therefore fair to affirm that 
the failure of these two approaches to 
reproduce the OES does not necessary mean that there is no pairing correlation 
in these nuclei. Two typical examples are $^{48}$Ca and $^{132}$Sn shown in 
Fig. \ref{fig:twopar_subplot}. These two nuclei show $\Delta^{(3)}_{C}$ values 
similar to neighboring nuclei below the closed shell. Indeed our calculations 
with the seniority model \cite{Xu2013247} show that the two-neutron spatial
correlations in these two nuclei are as strong as those in neighboring ones. 
Again, $\Delta^{(3)}_{C}$ is a good indication of the two-neutron correlation in 
these cases. Anyway, as mentioned above, comparisons between the theoretical gaps from even-even nuclei can only be compared with the empirical pairing gap in a semi-quantitative manner.

In summary, in this paper we compared the pairing gaps derived from four 
different OES formulas. We showed that $\Delta^ {(3)}_{C}$ 
gauge very well the nuclear pairing correlation since it removes to a large extent the 
contribution from the single-particle structure of the nuclear mean field and the shell effect. It can serve as a reliable filter if one is primarily interested in evaluating the pairing effect in structure model calculations. This is particularly interesting for the study of dripline nuclei since experimental data on those nuclei are usually scarce and show large fluctuations of shell structure.
In addition, $\Delta^ {(3)}_{C}$  is expected to be less contaminated by the Wigner effect for nuclei around $N=Z$. We
have also shown that the  
strength of $\Delta^ {(3)}_{C}$ can be a good semi-quantative indication for  the two-particle spatial
correlation. This is supported by our calculations 
with the HFB model with Skyrme force as well as with the multi-shell seniority 
model. Moreover, we found that the weakening of $\Delta^ {(3)}_C(N)$ in some
neutron-rich nuclei indicates that the di-neutron correlation is weak in 
those nuclei. As examples, the pairing gaps and di-neutron spatial correlations in neutron-rich calcium and tin isotopes are evaluated in detail. 
Calculations for the different pairing gaps and OES with different functional quantatively agree well with each other as well as with experimental data. However, these results need to
be carefully examined when one approaches the dripline, where large deviations between different functionals and different definations of the pairing gaps may be seen.

\section*{Acknowledgment}
We thank R. Liotta for stimulating discussions and his reading of the manuscript.
This work was supported by the Swedish Research Council (VR) under grant Nos. 621-2012-3805, and
621-2013-4323.
The
calculations were performed on resources
provided by the Swedish National Infrastructure for Computing (SNIC)
at NSC in Link\"oping and PDC at KTH, Stockholm.

\end{document}